\documentclass[twocolumn,lineno,dvipsnames, adjustbox,usenames]{aastex63}
\usepackage{amssymb,amsmath,verbatim,mathtools,needspace,enumitem,etoolbox,graphicx,physics,microtype,ulem}
\usepackage{xspace}
\usepackage{color}
\usepackage{graphicx}
\usepackage{float}
\usepackage{amsmath}
\usepackage{acronym}
\usepackage{multirow}
\usepackage{svn-multi}
\usepackage{import}
\usepackage{hyperref}
\usepackage[normalem]{ulem}
\usepackage[tableposition=top]{caption}
\usepackage{xr}
\usepackage{aas_macros}

\newcommand{\msun}{\ensuremath{\mathrm{M}_\odot}}
\newcommand{\si}{\ensuremath{\sim}}
\newcommand{\beq}{\begin{equation}}
\newcommand{\eeq}{\end{equation}}
\newcommand{\beqa}{\begin{eqnarray}}
\newcommand{\eeqa}{\end{eqnarray}}
\newcommand{\ud}{\ensuremath{\mathrm{d}}}
\newcommand{\nform}{\ensuremath{N_{\mathrm{form}}}\xspace}

\newcommand{\comv}{\ensuremath{V_{\mathrm{C}}}\xspace}

\newcommand{\myr}{\ensuremath{\mathrm{Myr}}\xspace}
\newcommand{\gyr}{\ensuremath{\mathrm{Gyr}}\xspace}

\DeclareMathOperator{\cov}{Cov}

\newcommand{\rateunits}{\ensuremath{\mathrm{Gpc}^{-3} \mathrm{yr}^{-1}}\xspace}
\newcommand{\psiunit}{\ensuremath{\mathrm{M}_{\odot}~\mathrm{Mpc}^{-3}\mathrm{yr}^{-1}\xspace}}

\newcommand{\TheEvent}{GW150914}

\newcommand{\sv}[1]{{\color{black}{\sf{#1}}}}

\newcommand{\tauMDthirtyK}{\ensuremath{0.93^{+0.35}_{-0.31}}}
\newcommand{\alphaMDthirtyK}{\ensuremath{2.54^{+0.68}_{-0.60}}}
\newcommand{\CMDthirtyK}{\ensuremath{2.90^{+0.23}_{-0.27}}}
\newcommand{\betaMDthirtyK}{\ensuremath{5.57^{+0.59}_{-0.54}}}
\newcommand{\psiMDthirtyK}{\ensuremath{-1.72^{+0.56}_{-0.60}}}
\newcommand{\NoneMonth}{\ensuremath{33900}}
\newcommand{\thirtyKtime}{\ensuremath{27\,\mathrm{days}}}

\newcommand{\tauMDthirtyKPrompt}{\ensuremath{0.21^{+0.13}_{-0.11}}}
\newcommand{\alphaMDthirtyKPrompt}{\ensuremath{2.83^{+0.45}_{-0.49}}}
\newcommand{\CMDthirtyKPrompt}{\ensuremath{3.10^{+0.23}_{-0.24}}}
\newcommand{\betaMDthirtyKPrompt}{\ensuremath{5.62^{+0.41}_{-0.44}}}
\newcommand{\psiMDthirtyKPrompt}{\ensuremath{-2.08^{+0.56}_{-0.63}}}
\newcommand{\NoneMonthPrompt}{\ensuremath{56740}}
\newcommand{\thirtyKtimePrompt}{\ensuremath{16\,\mathrm{days}}}

\newcommand{\alphaMDthirtyKSlow}{\ensuremath{2.62^{+0.71}_{-0.77}}}
\newcommand{\CMDthirtyKSlow}{\ensuremath{3.03^{+0.44}_{-0.49}}}
\newcommand{\betaMDthirtyKSlow}{\ensuremath{5.67^{+0.65}_{-0.65}}}
\newcommand{\tauMDthirtyKSlow}{\ensuremath{9.46^{+3.60}_{-3.33}}}
\newcommand{\psiMDthirtyKSlow}{\ensuremath{-1.88^{+0.70}_{-0.65}}}
\newcommand{\NoneMonthSlow}{\ensuremath{2310}}
\newcommand{\thirtyKtimeSlow}{\ensuremath{13\,\mathrm{months}}}

\newcommand{\tauMDthirtyKLog}{\ensuremath{0.19^{+0.12}_{-0.09}}}
\newcommand{\alphaMDthirtyKLog}{\ensuremath{2.09^{+0.36}_{-0.38}}}
\newcommand{\CMDthirtyKLog}{\ensuremath{3.39^{+0.13}_{-0.14}}}
\newcommand{\betaMDthirtyKLog}{\ensuremath{5.09^{+0.40}_{-0.40}}}
\newcommand{\psiMDthirtyKLog}{\ensuremath{-2.18^{+0.49}_{-0.55}}}
\newcommand{\NoneMonthLog}{\ensuremath{25980}}
\newcommand{\thirtyKtimeLog}{\ensuremath{35\,\mathrm{days}}}

\newcommand{\vmr}{\ensuremath{\mathcal{R}_m}\xspace}
\newcommand{\vfr}{\ensuremath{\mathcal{R}_f}\xspace}

\newcommand{\dmr}{\ensuremath{{R}_m}\xspace}

\begin{document}

\title{Measuring the star formation rate with gravitational waves from binary black holes}
\author[0000-0003-2700-0767]{Salvatore Vitale}
\email{salvatore.vitale@ligo.org}
\affiliation{LIGO, Massachusetts Institute of Technology, Cambridge, Massachusetts 02139, USA}
\affiliation{Kavli Institute for Astrophysics and Space Research, Massachusetts Institute of Technology, Cambridge, Massachusetts 02139, USA}
\author[0000-0003-1540-8562]{Will M.~Farr}
\email{will.farr@stonybrook.edu}
\affiliation{Department of Physics and Astronomy, Stony Brook University, Stony Brook, NY, 11794, USA}
\affiliation{Center for Computational Astrophysics, Flatiron Institute, 162 Fifth Avenue, New York NY 10010, USA}
\affiliation{Birmingham Institute for Gravitational Wave Astronomy, University of Birmingham, Birmingham, B15 2TT, UK}
\author[0000-0003-3896-2259]{Ken~K.~Y.~Ng}
\affiliation{LIGO, Massachusetts Institute of Technology, Cambridge, Massachusetts 02139, USA}
\affiliation{Kavli Institute for Astrophysics and Space Research, Massachusetts Institute of Technology, Cambridge, Massachusetts 02139, USA}
\author[0000-0003-4175-8881]{Carl L.~Rodriguez}
\affiliation{Kavli Institute for Astrophysics and Space Research, Massachusetts Institute of Technology, Cambridge, Massachusetts 02139, USA}
\affiliation{Harvard Institute for Theory and Computation, 60 Garden St, Cambridge, MA 02138, USA}

\begin{abstract}
A measurement of the history of cosmic star formation is central to understand
the origin and evolution of galaxies. The measurement is extremely
challenging using electromagnetic radiation: significant modeling is required to
convert luminosity to mass, and to properly account for dust attenuation, for
example. Here we show how detections of gravitational waves from inspiraling
binary black holes made by proposed third-generation detectors can be used to
measure the star formation rate (SFR) of massive stars with high precision up to
redshifts of \si 10. 
Depending on the time-delay model, the predicted detection rates ranges from $\sim \NoneMonthSlow$ to $\sim \NoneMonthPrompt$ per month with the current measurement of local merger rate density.
With 30000 detections, parameters describing the volumetric
SFR can be constrained at the few percent level, and the
volumetric merger rate can be directly measured to 3\% at $z\sim 2$. Given a
parameterized SFR, the characteristic delay time between binary
formation and merger can be measured to $\sim 60\%$.
\end{abstract}

\section{introduction}
The binary black holes (BBHs) detected by the ground-based gravitational-wave (GW) detectors LIGO~\citep{Harry:2010zz} and Virgo~\citep{TheVirgo:2014hva} all merged in the local universe~\citep{2016PhRvL.116x1102A,GW151226-DETECTION,2016PhRvX...6d1015A,2017PhRvL.118v1101A,2017ApJ...851L..35A,2017PhRvL.119n1101A,2018arXiv180511579T}.
These detections have allowed to measure the \emph{local} merger rate of BBHs at {$[24.4-111.7]$}~\rateunits (90\% credible interval~\citep{o2rates}).
The sensitivity of advanced detectors limits to {z\si 1} the maximum redshift at which an heavy BBH, \sv{with total mass of about 60~\msun}, such as \TheEvent{} can be detected, whereas heavier systems, including intermediate mass black hole binaries, could be observed farther away~\citep{2017arXiv170908079C,2019arXiv190209485H,2016PhRvL.116x1102A,GW151226-DETECTION,2016PhRvX...6d1015A,2017PhRvL.118v1101A,2017ApJ...851L..35A,2017PhRvL.119n1101A,2018arXiv180511579T,2017PhRvD..96b2001A,o2rates}.

As the LIGO and Virgo instruments progress toward their design sensitivity~\citep{2016LRR....19....1A}, and the network of ground-based detectors grows, it will be possible to detect BBH at redshifts greater than 1 (the exact value depending on the BBH mass). This can potentially allow us to probe the merger rate of BBHs through a significant distance range, and check how it varies with redshift~\citep{2018arXiv180510270F}.

While this might provide precious information on the evolution of the merger rate, it would be interesting to access sources at even higher redshifts.
Since compact binaries are constituted of neutron stars and black holes, leftovers of main-sequence stars, a measurement of their abundance at different stages of cosmic history can potentially tell us something about the star formation rate (SFR).
This latter is currently measured using various electromagnetic probes (\cite{2013ApJ...770...57B,2014ARA&A..52..415M}). However, electromagnetic probes do not directly track the amount of matter being formed on a galaxy. Instead, they track the luminosity, which then is linked to the mass production through several steps of modeling (e.g. on the initial mass function). Furthermore, dust extinction can significantly reduce the bolometric luminosity of a galaxy, or alter its spectral content, which is a key ingredient to infer the SFR from light. These limitations are particularly severe at redshifts above 3 where, additionally, fewer data points are available from electromagnetic observations~\footnote{We notice that this might become less true as future telescope get online, in the timescale relevant for the realization of third-generation GW detectors.}
It would thus be valuable to have an independent way of measuring the star formation at high redshifts, possibly by directly tracking masses, rather than light. Gravitational-wave signals can be used to that goal, as they directly encode information about the mass of the source.
Two proposals for third-generation (3G) ground-based detectors are currently being pursued, which would allow to detect BBHs at large redshifts: the Einstein Telescope~\citep{2010CQGra..27s4002P} (ET) and Cosmic Explorer (CE)~\citep{2017CQGra..34d4001A}.
Using the local merger rate calculated by the LIGO and Virgo collaborations it has been estimated that $[1-40]\times10^4$ BBHs merge in the universe per year~\citep{2017PhRvL.118o1105R}. ~\cite{Vitale3G} have shown how BBH can be detected all the way to redshift of \si15 by networks of 3G detectors. Since that is a significant fraction of the volume of the universe, one would thus expect that a large fraction of merging BBH would be detectable. Indeed, \cite{2017PhRvL.118o1105R} estimate that 99.9\% of the BBH mergers will be detectable by 3G detectors~\footnote{In this Letter we solely focus on BBHs. Previous work exists for binary neutron stars~\citep{VanDenBroeck:2010vx,2019ApJ...878L..13S}.}.
In this Letter we show how, under quite generic hypotheses, accessing BBHs with 3G gravitational-wave detectors, allows for a direct inference of the merger rate and the SFR all the way to redshifts of $\sim 10$.

\section{Event rates}\label{sec:rates}

As sources are detected in a gravitational-wave detector network, one can
estimate their redshifts \citep{Vitale3G,2016ApJ...825..116F,2015PhRvD..91d2003V}
and measure their detection rate in the local frame.  Let~\footnote{We will use the subscript ``f'' for quantities related to the formation of binaries, and ``m'' for quantities related to their merger.}  $\dmr(z_m)\equiv \frac{\ud N_m}{\ud t_d \ud z}$ be the total redshift rate density of mergers in the detector frame (the number of mergers per detector time per redshift).
The shape of this function, given the uncertainty in the observed
redshift of the detected sources, can be inferred with hierarchical
analysis~\citep{Mandel:2010,Hogg:2010,Youdin:2011,Farr:2011}.

The redshift rate density can be written in terms of the volumetric
total merger rate in the source frame, $\vmr(z_m)\equiv \frac{\ud N_m}{\ud V_c \ud t_s}$ as

\begin{equation}
\label{Eq.DifferentialRate}
\dmr(z_m)  = \frac{1}{1+z_m}\frac{\ud V_c}{\ud z}\vmr(z_m),
\end{equation}

where the $1+z_m$ term arises from converting source-frame time to detector-frame time~\citep{Dominik:2013tma}.

The volumetric merger rate {in galactic fields} depends on the star formation rate, the metallicity, and the delay
between the formation of the binary black hole progenitors and their eventual
merger.  All the systems that merge at a lookback time $t_m$ (or, which is
equivalent, at a redshift $z_m=z(t_m)$) are systems that formed at $z_f>z_m$ (or
$t_f > t_m$). The delay time distribution, $p(t_m | t_f,  \lambda)$, is the
probability density that a system that formed at time $t_f$ will merge at time
$t_m$. This function may depend on an (unknown) time scale, the parameters of
the system that is merging, and possibly other parameters.  We capture this
dependence using parameters $\lambda$.

We can write the merger rate at redshift $z_m$ as a function of the black hole
binary volumetric formation rate, $\vfr\left(z_f \right)$:
\beqa\label{Eq.VolumetricRateRed}
\vmr(z_m) &=& \int_{z_m}^{\infty}{ \ud z_f \frac{\ud t_f}{\ud z_f}\vfr(z_f) p(t_m| t_f, \lambda)}
\eeqa
Here we assume that volumetric formation rate $\vfr(z_f)$ is simply proportional
to the star formation rate density at the same redshift, $\psi(z)$ (see Eq.~\eqref{Eq.MDSFR}) and to the efficiency $\eta(z)$:
\beq\label{Eq.RofF}
\vfr(z_f)\equiv \frac{\ud \nform}{\ud \comv \ud t_f} \propto \eta(z_f) \psi(z_f).
\eeq
The fact that the merger rate is proportional to the star formation rate \emph{at the same redshift} is a reasonable assumption~\citep{GW150914-STOCHASTIC,2014ARA&A..52..415M}, since the life-time of massive stars that will become black holes is of the order of tens of Myr and hence negligible when compared to the other time-scales of interest. 
The efficiency $\eta(z)$ takes into account the fact that of the star formation at a given redshift, only the fraction $\eta(z)$ with low metallicity will result in heavy black hole formation. Following \cite{2016Natur.534..512B} we define $\eta(z)$ as the fraction of star formation that has metallicity below 10\% of the solar metallicity, and calculate it as:
\beq\label{Eq.Efficiency}
\eta(z)\equiv \int_{-\infty}^{\log(0.1 Z_\odot)}{\Phi(\log Z_{\mathrm{mean}}(z),0.5) \,\mathrm{d}Z}\;,
\eeq
where $\Phi$ is the cumulative distribution function of the metallicity at redshift $z$, assumed to be a Gaussian distribution with mean $Z_{\mathrm{mean}}$, and 0.5 dex of uncertainty, Eq.~\eqref{Eq.VolumetricRateRed} of \cite{2016Natur.534..512B}.

We do not account here for
eventual contributions to the formation rate arising from binaries that do not
form in galactic fields (e.g.\ binaries from globular clusters or from population
III stars). The methods we use can be extended to account for multiple formation
channels; we discuss this possibility further below.

Both the formation rate and the time delay distribution might depend on some
intrinsic  properties of the of the binary being formed, e.g., the component
masses~\citep{Dominik:2013tma}. These dependencies can be included in an extension of our analysis in a
straightforward manner, by adding the masses and other parameters to $\lambda$
and marginalizing them in Eq.~\eqref{Eq.VolumetricRateRed}. However,
for this proof-of-principle study we will assume these details can be neglected.

In this work we will follow two different approaches. First, we will assume
that nothing is known about the true functional form of the SFR and the
time-delay distribution. In this case, we use a non-parameteric Gaussian process
algorithm to directly measure the volumetric rate density in the detector frame,
$\vmr(z)$. Next, we will show that assuming the parameterized functional
form of both the SFR and the time-delay distribution, the parameters on which they depend can be measured from the GW
detections.

\section{Simulated signals}\label{sec:method}

\begin{figure}
\vskip 0.2cm
\includegraphics[width=\columnwidth]{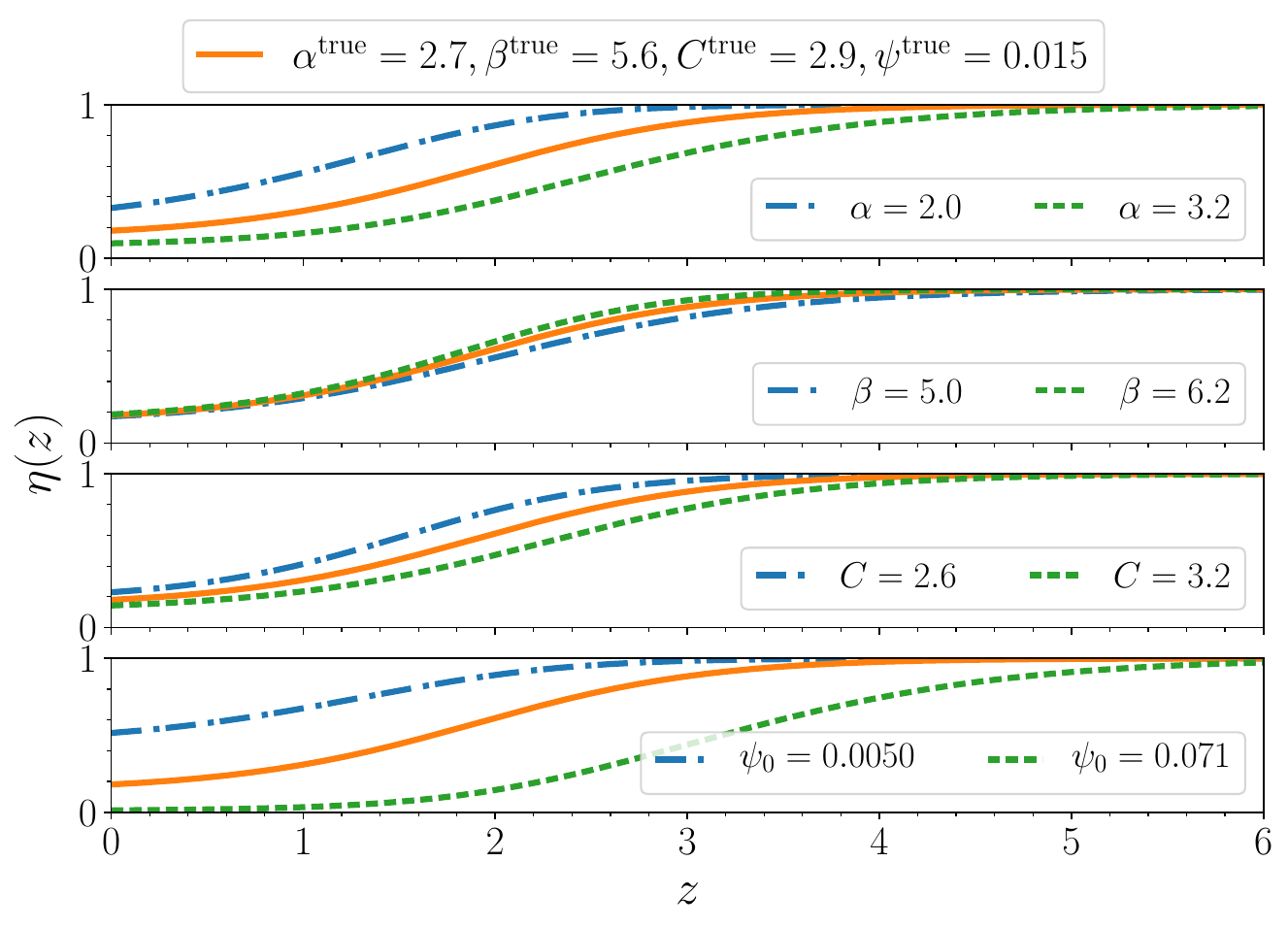}
\caption{Efficiency $\eta(z)$ plotted as function of the redshift. In each panel, the orange curve is obtained using the nominal MD SFR. The other curves are obtained by varying, in turn, each of the 4 parameters controlling the MD SFR. The range of variation is taken to be representative of the uncertanties we find in Sec.~\ref{Sec.Results}.} \label{Fig.Efficiency}
\end{figure}

To demonstrate how the cosmic BBH merger rate can be measured, we generate 30,000 synthetic BBH detections in each time-delay model with realistic redshift uncertainty (see below)
\citep{Vitale3G}. We assume that the SFR is the Madau-Dickinson (MD)
star-formation rate, which can be written:
\begin{eqnarray}\label{Eq.MDSFR}
\psi_{MD}(z)&=&\psi_0 \frac{(1 + z)^{\alpha}}{1 + \left(\frac{1+z}{C}\right)^{\beta}},
\end{eqnarray}
with parameters $\alpha=2.7$, $\beta=5.6$, $C=2.9$ and $\psi_0 = 0.015\,\psiunit$
\citep{2014ARA&A..52..415M}.
The proportionality coefficient in Eq.~\eqref{Eq.RofF} is chosen such that the local BBH merger rate  $R_m(0)$ is equal to $50~\rateunits$, consistent with LIGO and Virgo's measurements. We notice that, since the SFR also affects the mean metallicity, and hence the efficiency $\eta(z)$~\citep{Belczynski2016}, the local merger rate is not simply proportional to $\psi_0$.
This is shown if Fig.~\ref{Fig.Efficiency}. Different panels show how $\eta(z)$ varies when the parameters describing the MD SFR are varied, one at the time. The range of variability is taken to be representative of the uncertainties we find in their measurement
in Sec.~\ref{Sec.Results}. We see that $\psi_0$ and $\alpha$ have similar effects in the efficiency, and we should thus expect them to be anti-correlated. C and $\beta$ have instead a milder effect on $\eta(z)$.

We consider two different functional
forms for the distribution of time-delays between formation and merger: an
exponential function with time scale parameter $\tau$:
\beq\label{Eq.TimeDelayExp}
p(t_m | t_f ,\tau) = \frac{1}{\tau} \exp{\left[-\frac{\left(t_{f} - t_{m}\right)}{\tau}\right]}
\eeq
and a distribution uniform in the logarithm of the time delay:

\beq
 p(\log(t_m-t_f)) \propto \left\{
\begin{array}{ll}
      1 & \;10 \myr < t_m-t_f<10 \gyr \\
      0 & \;\mathrm{otherwise} \\
\end{array}\label{Eq.FlatLog}
\right.
\eeq
The true redshifts of the sources under both delay assumptions are randomly
drawn from Eq.~\eqref{Eq.DifferentialRate}, after normalizing it to unity in the
redshift range $z\in [0,15]$.

\begin{figure}[htb]
\includegraphics[width=\columnwidth,clip=true]{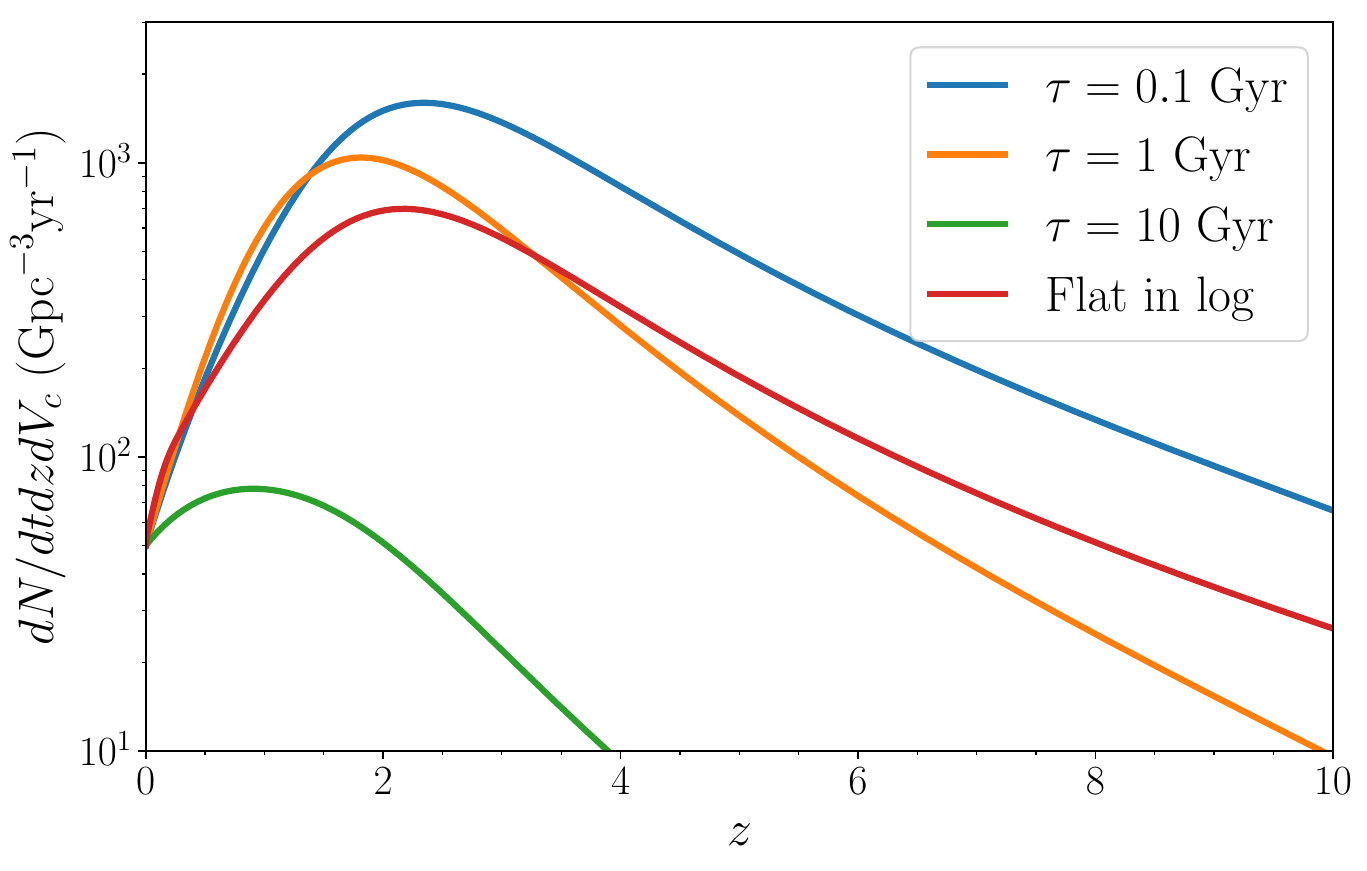}
\caption{The merger redshift distribution of the simulated population of BBH. We assume
a Madau-Dickinson SFR, and four different prescriptions for the time delay
between formation and merger: an exponential time delay with e-fold time of
$100\myr$, $1\gyr$ and $10\gyr$, and a uniform-in-log distribution, with a
minimum of 10\myr and a maximum of 10\gyr.}\label{Fig.InjectedPz}
\end{figure}

In Fig.~\ref{Fig.InjectedPz} we show the redshift distribution of the simulated
BBH merger events using the exponential time delay with $\tau = 0.1 \, \gyr$, $1 \,
\gyr$, $10\, \gyr$~\footnote{These three values, as well as the minimum and maximum time delay in Eq.~\eqref{Eq.FlatLog} are chosen to cover a reasonable range of characteristic time delays~\citep{Dominik:2012kk,Nakar:2007yr,Berger:2006ik,Belczynski:2006br,Belczynski:2001uc,2013ApJ...779...72D}}, and with the flat-in-log distribution at a fixed local merger rate density of 50~$\rateunits$.
The estimated number of events in one month is $M=\NoneMonthPrompt,\NoneMonth,\NoneMonthSlow$ and $\NoneMonthLog$ respectively.
The corresponding time to observe 30000 events is $T=\thirtyKtimePrompt,\thirtyKtime,\thirtyKtimeSlow$ and $\thirtyKtimeLog$ respectively.

The redshift of detected BBH cannot be perfectly measured using GW detectors. We
approximate the results of a full analysis of a three-detectors 3G network
\citep{Vitale3G} by assuming that the likelihood function for the true redshift follows a log-normal distribution conditioned on the true redshift with
standard deviation 
$\sigma_{\mathrm{LN}} (z_{\mathrm{true}})=0.017 z_{\mathrm{true}}+0.012$.

We do not explicitly draw mass values or calculate a signal-to-noise ratio. As
long as one works with BBH of total mass above \si15\msun, all sources are
detectable by 3G networks including the CE up to redshifts were the merger rate
becomes negligible~\citep{Vitale3G,2017PhRvL.118o1105R}.

Once the catalog of simulated events and the corresponding redshift likelihoods
have been generated, our analysis proceeds hierarchically~
\citep{Mandel:2010,Hogg:2010,Youdin:2011,Farr:2011}.  We assume that the
production of gravitational-wave sources is an (inhomogeneous) Poisson process,
with rate density 

\beq
 \vmr \left( z \mid \lambda \right),\nonumber
\eeq

depending on some parameters $\lambda$.  Therefore the posterior for the
population-level parameters given (synthetic) data for 30000 events, $\vec{d}\equiv \left\{ d_i \right\}_{i=1}^{M}$ is~
\citep{Foreman-Mackey:2014,Farr:2015,Youdin:2011}
\begin{multline}\label{Eq.PofLambda}
p\left( \lambda \mid \vec{d}\,\right) \propto \left[\prod_{i=1}^{M} \int \dd z_i \, p\left( d_i \mid z_i \right) \dmr\left(z_i \mid \lambda \right)\right]  e^{- \chi} \,p\left(\lambda\right) \\ \simeq \left[\prod_{i=1}^{M} \frac{1}{M_i} \sum_{j=1}^{M_i} \dmr \left( z_{ij} \mid \lambda\right)\right] e^{- \chi} \, p\left(\lambda\right),
\end{multline}

where $\chi\equiv \int \dd z \, \dd t_d \, \dmr\left(z \mid \lambda\right)$, $z_i$ is the redshift of event $i$; $p\left( \lambda \right)$ is a prior imposed on the parameters describing the merger rate density; and we use $M_i$
samples, $\left\{z_{ij}\right\}_{j=1}^{M_i}$, drawn from a density proportional
to the likelihood, $z_{ij} \sim p\left( d_i \mid z_{ij} \right) \dd z_{ij}$, to approximate the marginalisation integral over $z_i$.

\section{Results}\label{Sec.Results}

We desire to understand how well we can expect to constrain the merger rate
density and the time delay distribution from our synthetic data set of 30000 observations.

We first consider an unmodeled approach, where nothing is assumed about the
underlying SFR function and time-delay distribution other than that it is
relatively smooth~\citep{Foreman-Mackey:2014}.  We assume that the log of the
merger rate can be described by a piecewise-constant function over $K = 29$
redshift bins. %
To ensure there are enough samples in each bin, we choose the bins in the following way:  $0\leq z<0.32$ for the first bin, while the remaining bins are uniformly distributed in $\log (1+z)$ with $z\in[0.32,15)$ so that the log of merger rate is
\begin{equation}
\log \vmr
= \begin{cases}
n_1 & 0 \leq z < z_1 \\
\ldots & \\
n_i & z_{i-1} \leq z < z_i \\
\ldots & \\
n_K & z_{K-1} \leq z < z_K
\end{cases},
\end{equation}
and we treat the per-bin merger rates, $n_i$, as parameters, $\lambda$, in
Eq.~\eqref{Eq.PofLambda}.  
We apply a squared-exponential Gaussian Process prior on the $n_i$, which has a covariance kernel of
\begin{equation}
\cov\left( n_i, n_j \right) = \sigma^2 \exp\left[-\frac{1}{2}\left(\frac{z_{i-1/2} - z_{j-1/2}}{l}\right)^2\right],
\end{equation}
with $z_{i-1/2} = \left( z_{i} - z_{i-1} \right)/2$ the midpoint of the $i$th
redshift bin.  We treat the variance of the $n_i$, $\sigma^2$, and the
correlation length in redshift space, $l$, as additional parameters in the fit.
The squared-exponential Gaussian Process prior enforces the smoothness of the
merger rate on scales that are comparable to or larger than $l$ (which may be
much larger than the bin spacing if the data support it), and guards against
over-fitting when $K$ is large~\citep{Foreman-Mackey:2014}.

\begin{figure}[htb]
\includegraphics[width=\columnwidth,clip=true]{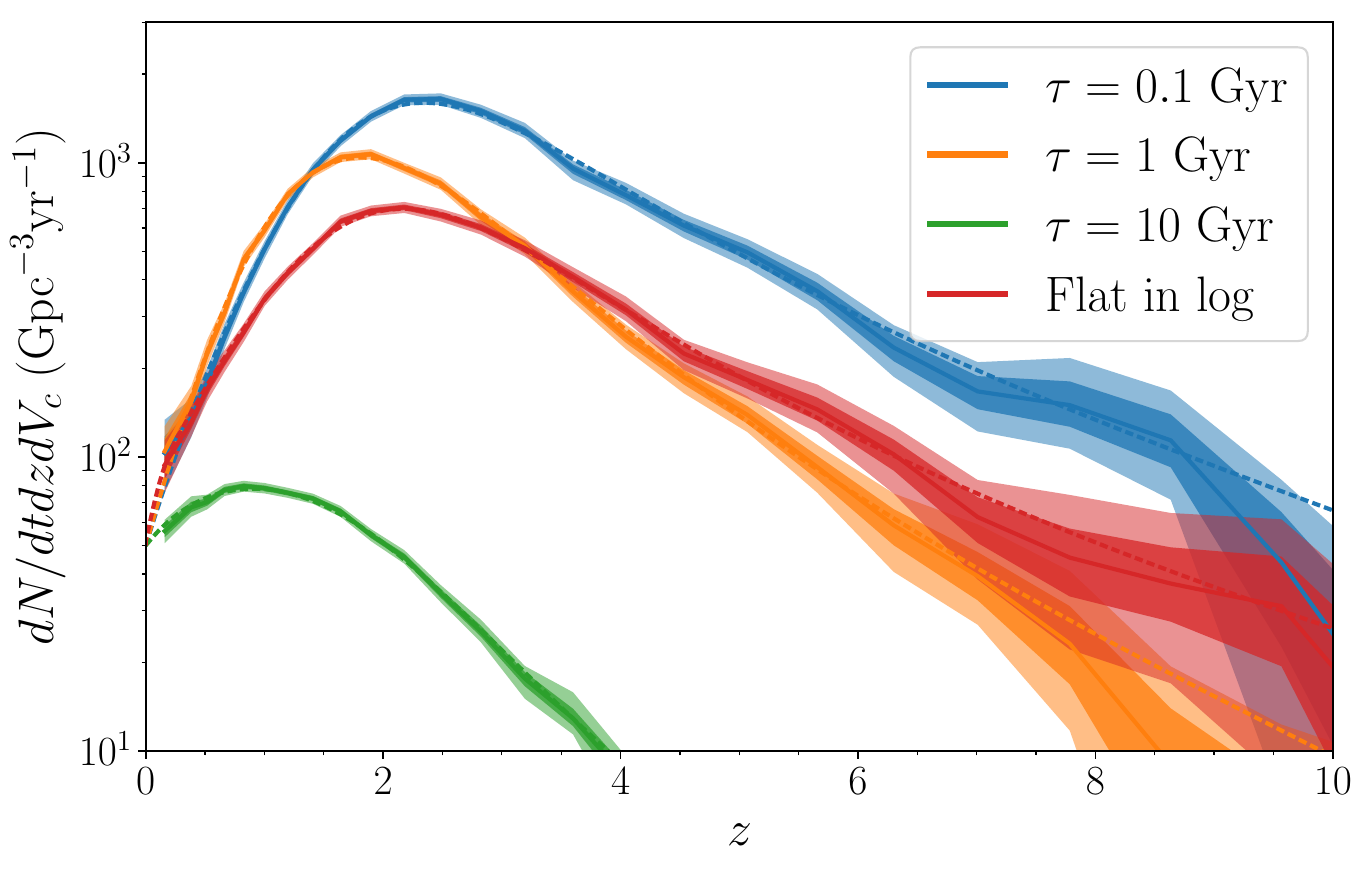}
\caption{Posterior on the volumetric merger rate density calculated using an unmodeled
approach. The dashed lines are the true rates under the four possible time delay
distributions we consider. Full lines give the median measurement, while the
bands report the 68\% and 95\% credible intervals. Near the peak $z\sim 2$ the
uncertainty in the rate estimate is $\sim 3\%$ for $\tau=0.1\gyr,1\gyr$ and flat-in-log models.
The uncertainty rises to $10\%$ in $\tau=10\gyr$ model around the peak $z\sim 1$, as the total number of events is 10 times smaller than the numbers in other models.
The small
systematic offset for the flat-in-log and prompt data sets is likely due to a
$100 \, \mathrm{Myr}$ lower limit on the delay time imposed for numerical
stability; see the corresponding discussion in the parameterized model
results.}\label{Fig.MeasuredGP}
\end{figure}

The results for this fit are shown in Fig.~\ref{Fig.MeasuredGP}, where for each
true synthetic population we show the median posterior on the piecewise-constant
$\dd N / \dd V_c \dd t_d$, together with 68\% and 95\% (1- and 2-sigma) credible
intervals. We see that the unmodeled GP method pinpoints the merger rates so
precisely that all four distributions are clearly distinguishable; near $z \sim
2$ the uncertainty in the measured merger rate is $\sim 3\%$. At moderate
redshifts, $z<4$, the uncertainties are smaller than the separation between
different populations.  At larger redshifts the measurement becomes more
uncertain, and overlaps exist. This is due to a combination of two effects: from
one side, fewer sources merge, and hence are detected, at those redshifts; from
the other, the uncertainty in their measured redshift is higher.  The advantage
of this approach over a more rigid parameterization of the merger rate is that
it can fit \emph{any} sufficiently smooth merger rate; a disadvantage is that we
learn nothing individually about the time-delay distribution or the star
formation rate, since it they are completely degenerate in this flexible model.

Next, we want to verify how well we can measure the characteristic parameters of the SFR and time-delay distribution \emph{assuming} we know their functional forms.

For this analysis, we take the MD SFR and the exponential time-delay
distribution as models, treating the parameters $\alpha$, $\beta$, $C$, $\psi_0$ as well
as the time-delay scale $\tau$ as unknowns. We then calculate the posterior for
$\lambda_{MD}=\{\alpha,\beta,C,\psi_0,\tau\}$ with Eq.~\eqref{Eq.PofLambda}.  Note that
the parameterized model with an exponential time delay cannot perfectly match the flat-in-log data-generating
model, no matter what value of $\tau$ is used.

\begin{figure}
\includegraphics[width=\columnwidth,clip=true]{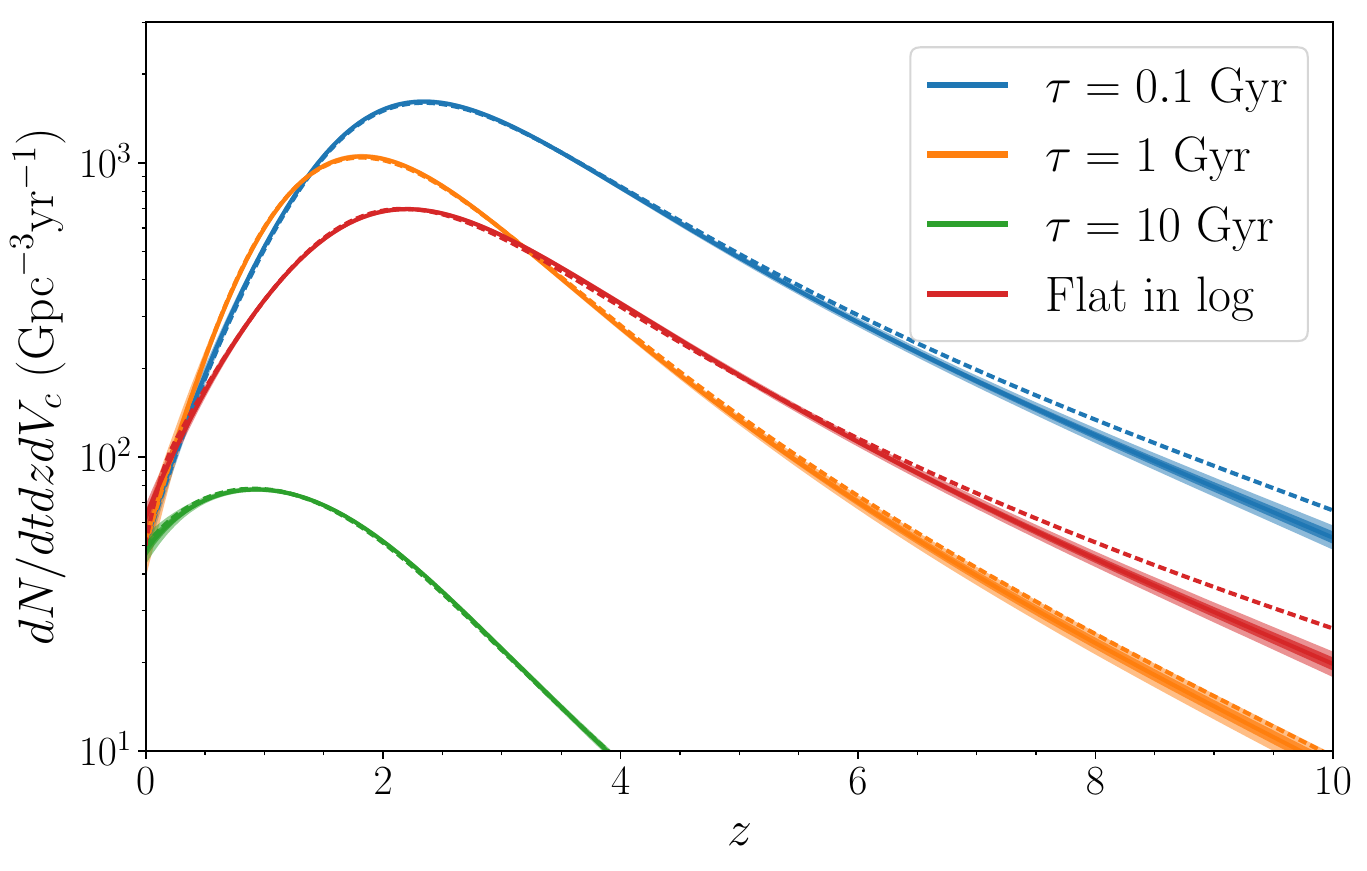}
\caption{Posterior on the merger rate density calculated from the parameterized
fits described in the text.  Dashed lines show the true merger rate
distributions for our models.  Solid lines give the posterior median and dark
and light bands the 68\% and 95\% credible intervals.  See the text for more details}
\label{fig:dNdz-parameteric} \end{figure}

\begin{figure}
\vskip 0.2cm
\includegraphics[width=\columnwidth]{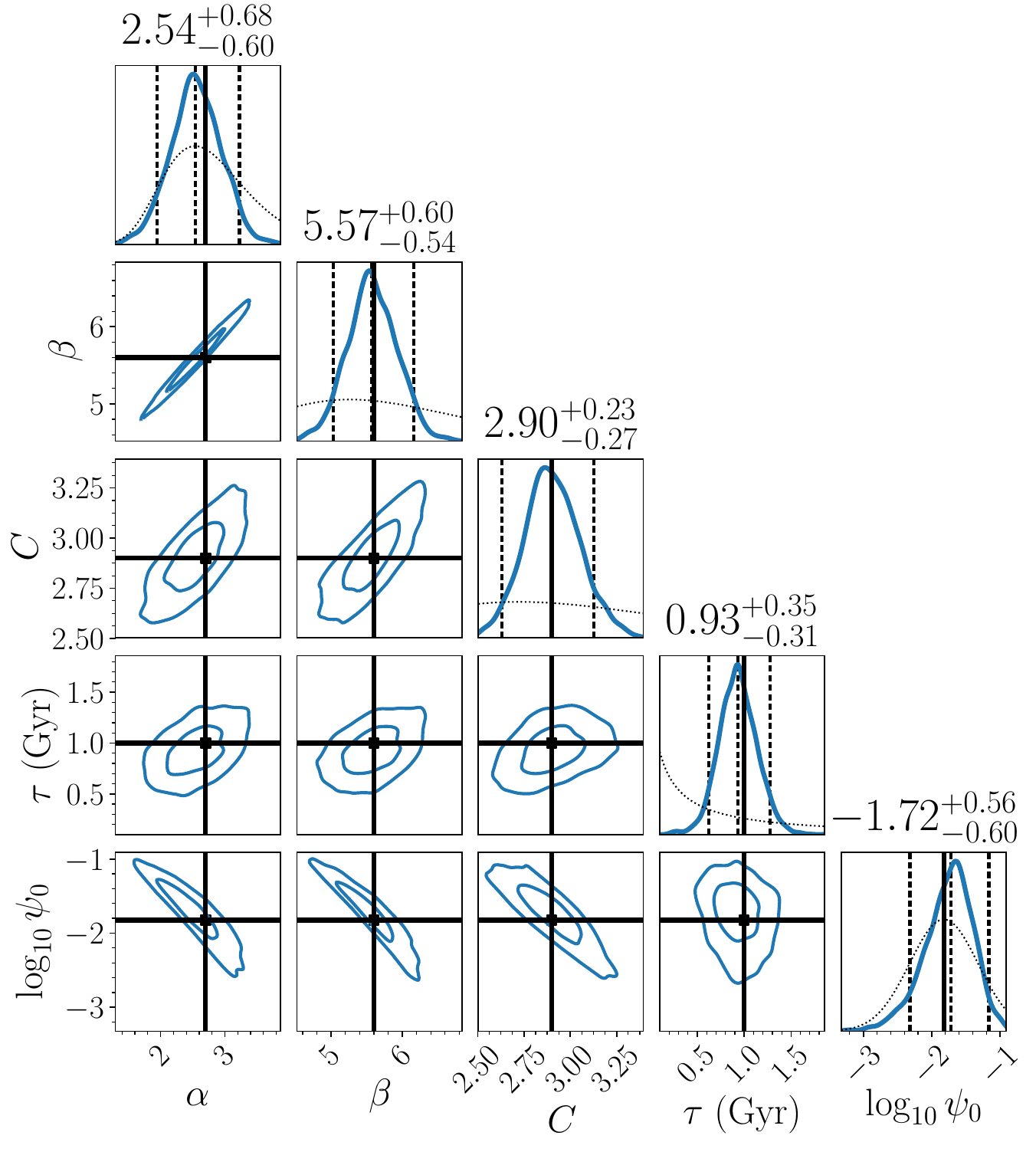}
\caption{The posterior distribution for the time-delay timescale and the MD SFR
parameters after 30000  detections in the $1 \, \mathrm{Gyr}$ delay timescale
scenario.
Truth is indicated by blue lines.
$\psi_0$ is in the unit of \psiunit.
Dashed lines indicate the highest posterior density 90\% credible interval; star formation rate parameters are
measured to few percent precision, and the delay timescale is measured to $\sim
60\%$.  Plot labels give the median and the highest posterior density 90\% credible interval for each
parameter.} \label{Fig.MDPosteriorsAll}
\end{figure}

We use log-normal priors with a width of $\simeq 0.25$ in
the log for $\alpha , \beta$ and $C$, reflecting an approximation to the uncertainty in the determination of
the SFR~\citep{2014ARA&A..52..415M}.
We also use a log-normal prior for $\psi_0$, with a prior large enough that the posterior is not truncated.
For $\tau$, we use a width of 2 in log to cover the whole dynamical range from 0.1~\gyr to 10~\gyr.
The uncertainties are large enough that the
posterior distributions are not truncated by the prior; with 30000 simulated detections we obtain meaningful constraints on the SFR parameters at the few
percent level and the time delay at a few tens of percent in all models.
We place a lower bound on the time-delay parameter $\tau \geq 100 \, \mathrm{Myr}$
in order to ensure numerical stability in our computation of the integral in
Eq.~\eqref{Eq.VolumetricRateRed}.  This results in some discrepancy between the
fit and the data-generating distribution for the ``prompt'' data set; the prompt
data is recovered in the limit $\tau \to 0$, but as this is excluded by our
prior there is a bias in the fit, particularly at high redshift where timescales
of $100 \, \mathrm{Myr}$ are a significant fraction of the age of the universe.
The inferred posterior on the merger rate redshift density is shown in Figure
\ref{fig:dNdz-parameteric}.  In Fig.~\ref{Fig.MDPosteriorsAll} we show
posteriors for the parameters $\lambda_{MD}$ for the set of events with
$\tau=1\,\gyr$.

After 30000 detections in the $1 \, \mathrm{Gyr}$ scenario, the scale
factor of the time delay distribution can be measured with relative uncertainty
of 60\% {(at 90\% credible interval)}: $\tau=\tauMDthirtyK$.
The parameters controlling the peak and high-$z$ slope of the MD SFR can also be measured with precision of $\sim 20\%$ or better, and we obtain $\beta=\betaMDthirtyK$, and $C=\CMDthirtyK$. On the other hand, $\alpha=\alphaMDthirtyK$ and $\log_{10}\psi_0=\psiMDthirtyK$ are only marginally narrower than their priors.

Many correlations are visible in Fig.~\ref{Fig.MDPosteriorsAll}, which is worth discussing, as they arise from different astrophysical factors.
First, $\tau$ and $C$ show a clear correlation, which can be understood as follows.
If $C$ increases then the peak of the SFR moves to higher redshift. In order to keep the \emph{observed} merger rate fixed the delay time must increase.
On the other hand, $\psi_0$ and $\alpha$ are anti-correlated as they both affect the efficiency $\eta(z)$, and hence the merger rate, in a similar way.
When $\psi_0$ increases, the total star formation at each redshift, as well as the metallicity, increase, which reduces the overall efficiency, bottom panel of Fig.~\ref{Fig.Efficiency}.
To compensate the loss of efficiency, $\alpha$ (and to a smaller extent, $C$) need to be decrease, as shown by the blue dot-dashed curves in Fig.~\ref{Fig.Efficiency}. This explains the anti-correlations seen for the pairs  $(\alpha,\psi_0)$ and $(C,\psi_0)$ in Fig.~\ref{Fig.MDPosteriorsAll}. Finally, $\beta$ does not affect much $\eta(z)$, Fig.~\ref{Fig.Efficiency}, and the correlation seen in Fig.~\ref{Fig.MDPosteriorsAll} for the pair $(\beta,\psi_0)$ is really only a consequence of the fact that $\alpha$ and $\beta$ are strongly correlated.

The parameter recovery for the other scenarios is similar; but for the flat in log scenario the systematic bias from model mismatch is significantly larger the statistical uncertainty.
The parameter estimates obtained from all scenarios are given in Table\ \ref{Tab.NearbyInj}.
Determination of the time delay distribution and the parameters of the star formation rate also allow measurement of the total number of BBH mergers per solar mass of star formation (not shown).

\begin{widetext}
\begin{table*}[htb]
\centering
\caption{Median and {90\%} credible intervals for the posterior of the MD and time-delay scale. The first column reports which event set is used. \label{Tab.NearbyInj}}
\begin{tabular}{c c c c c c}
\hline
\hline
True time-delay & $\alpha$ & $\beta$     &C& $\tau\,(\gyr)$ & $\log_{10}\psi_0$\\ \hline
Exp. $\tau=0.1\,\gyr$  &   \alphaMDthirtyKPrompt&    \betaMDthirtyKPrompt&  \CMDthirtyKPrompt & \tauMDthirtyKPrompt & \psiMDthirtyKPrompt   \\
Exp. $\tau=1.0\,\gyr$  &   \alphaMDthirtyK&    \betaMDthirtyK&  \CMDthirtyK & \tauMDthirtyK & \psiMDthirtyK  \\
Exp. $\tau=10\,\gyr$  &   \alphaMDthirtyKSlow&    \betaMDthirtyKSlow&  \CMDthirtyKSlow & \tauMDthirtyKSlow & \psiMDthirtyKSlow  \\
Flat Log  &   \alphaMDthirtyKLog&    \betaMDthirtyKLog&  \CMDthirtyKLog & \tauMDthirtyKLog & \psiMDthirtyKLog\\ \hline
\end{tabular}
\end{table*}
\end{widetext}

\section{Discussion and outlook}
\label{sec:discussion}

In this Letter we have shown how next-generation ground-based detectors will enable using gravitational waves from binary black hole to infer their merger rate throughout cosmic history, even in absence any model for the star formation history.
On the other hand, if a modeled template is available for the star formation rate and for the time-delay distribution between formation and merger, we have shown how their characteristic parameters can be measured with 30000 simulated signals.

We have simulated four different ``Universes'', assuming the BBH formation rate is proportional to the Madau-Dickinson star formation rate. The coefficient of proportionality is a redshift- and SFR-dependent function that accounts for the fraction of SFR with metallicity below 10\% of the solar metallicity~\citep{Belczynski:2006br}.
The four data sets use four different prescriptions for the delay between formation and merger: flat in the logarithm of the time-delay, or exponential, with e-fold time of 0.1, 1 or 10~\gyr. 

The unmodeled approach yields a direct measurement of the volumetric merger rate $\vmr\equiv \ud N/\ud V_c\ud t_d$. Fig~\ref{Fig.MeasuredGP} shows the measurement obtained with 30000 simulated signals. The four models are clearly distinguishable, and have uncertainties much smaller than their separation for redshifts below $\sim 6$.
At larger redshifts, the uncertainties increase due to the smaller number of sources, and the larger uncertainty on their redshifts.

Including a model for the star-formation history and the time-delay distribution dramatically increases the power of the method, and the expense of its generality.
Using the Madau-Dickinson SFR, Eq.~\eqref{Eq.MDSFR}  and an exponential time-delay distribution with unknown e-fold time $\tau$ as templates, we have shown how all unknowns can be measured with good precision after 30000 simulated signals.
The measurement of the SFR parameters is not accurate for the universe with flat-in-log time delays, as one would have expected given the mismatch between the time-delay template and the actual time-delay distribution.
This kind of issues can be mitigated using templates with more parameters. The number of parameters will increase the computational cost of the analysis, and the uncertainty in the measurement. However, the number of detectable BBH is in the hundreds of thousand per year, which will compensate for the extra complexity of the model.

In this work we have made a few simplifying assumptions to keep the computational cost under control.
First, we have assumed that the time-delay distribution is the same for all sources at all redshifts, while in reality it will depend on the redshift of the source through the metallicity of the environment~\citep{2019MNRAS.482.5012C}. This limitation can be lifted, introducing a functional form that relates time delay to redshift and possible other parameters, that will eventually be marginalized over.
Relatedly, we have neglected the dependence of the SFR and time-delay distribution on the mass and spins of the sources. This is not an intrinsic limitation of the method, and can be easily folded in the analysis. As these extra parameters are accounted for, we would expect that more sources will be required to achieve the same precision. But, as mentioned above, in this work we have considered 30000 simulated signals, which correspond to a few weeks to one year of observing time, depending on the actual time-delay distribution. More detections will be available for these tests, hence compensating for the increased complexity of the model.

Finally, while generating the simulated signals, we have assumed that all sources come from galactic fields. There is growing evidence that at least a fraction of BBH detected by LIGO and Virgo have been formed in globular clusters~\citep{2015PhRvL.115e1101R,2016PhRvD..93h4029R}.
These sources would show a very different evolution with redshift, with a peak of the merger rate at higher redshift.
If black holes from Population III stars merge, they could also contribute to the total merger rate, probably with a peak above $z\sim 10$~\citep{Belczynski2016,2016MNRAS.456.1093K}.
Depending on the relative abundance of mergers in these channels, one could be able to calculate their branching ratios as a function of redshift. This would give information which is complementary to what can be obtained studying the mass, spin, and eccentricity distribution of gravitational-wave detections. The method we developed can be extended to account for multiple population, which we will explore in a future publication.

\section{Acknowledgments}

The authors would like to thank H.-Y.~Chen, M.~Fishbach, R.~O'Shaughnessy, C.~Pankow, T.~Regimbau, for useful comments and suggestions. 
We thank the anonymous referee for their useful comments.
SV acknowledges support of the National Science Foundation through the NSF award PHY-1836814.
SV and KKYN acknowledge the support of the National Science Foundation and the LIGO Laboratory.
LIGO was constructed by the California Institute of Technology and Massachusetts Institute of Technology with funding from the National Science Foundation and operates under cooperative agreement PHY-1764464.
The author would like to acknowledge the LIGO Data Grid clusters, without which the simulations could not have been performed.
This is LIGO document number P1800219.
\bibliography{draft}
\end{document}